\documentclass{PoS}

\title{Theory overview on amplitude analyses with charm decays}

\ShortTitle{Theory overview on amplitude analyses with charm decays}

\author{\speaker{Benoit Loiseau} 
\\
        Sorbonne Universit\'es, Universit\'e Pierre et Marie Curie, Sorbonne Paris Cit\'e, Universit\'e
 Paris Diderot, et IN2P3-CNRS, UMR 7585, Laboratoire de Physique Nucl\'eaire et de Hautes \'Energies,  
4 place Jussieu, 75252 Paris, France\\
        E-mail: \email{loiseau@lpnhe.in2p3.fr}}

\abstract{This contribution about amplitude analyses in multibody hadronic charm decays deals with some attempts to introduce theoretical constraints. 
Different effective hadronic  formalism approaches are mentioned. 
A recent work, based on a basic weak interaction process and a Chiral unitary model to account for the final state interaction, is described in details for the  $f_0(980)$ production in $D_s^+ \to \pi^+ \pi^+ \pi^-$ and $D_s^+ \to \pi^+ K^+ K^- $ decays.
Within the framework of the diagrammatic approach and flavor symmetry, a global analysis of two-body $D$ decays into a vector meson and a pseudoscalar meson is presented.  
A quasi-two-body QCD factorization model for $D$ decays into three mesons and its recent application to $D^0 \to K_S^0 \pi^+ \pi^-$ is outlined.
For processes with final-state pions and kaons and as an alternative to the sum of Breit-Wigner amplitudes, often used in experimental Dalitz-plot analyses, 
amplitude parametrizations, in term of unitary $\pi \pi$, $\pi K$ and $K \bar K$ form factors, are proposed. 
These parametrizations are derived from quasi-two-body factorization models.
}

\FullConference{VIII International Workshop On Charm Physics\\
		5-9 September, 2016\\
		Bologna, Italy}

\def\sf{s_\phi}

\def\ol{\overline}

\begin{document}

\section{Introduction}
\label{introduction}

There is an impressive set of hadronic multibody decay data for $D^0$, $D^+$ and $D_s^+$ decays~\cite{LHCbWS, GersabeckEO, BriereQCC}.
The Dalitz plots are characterized by an accumulation of events displaying the presence of meson resonances and their interferences~\cite{Rademaker16, AlbertoAMPAN, BennettAA, EvansLHCB, YuLu}.
The Standard model (SM) predicts null $CP$ asymmetries and some deviation could be a signal of physics beyond SM~\cite{PaulND, BucellaPRD88}.
Furthermore the study of $D^0$-$\bar D^0$ mixing might indicate the presence of new physics contributions~\cite{WilliamsCPLHCb, LenzTO, CiuchiniDDBM, CKMfit}.
Multibody hadronic decays of $D_{(s)}$ mesons consist of a weak process,  microscopic quark flavor changing process like $c \to d$ or $c \to s$ via the $W$ meson interaction, followed by hadronization and  final state meson-meson strong interaction processes.
Basic amplitude analyses are usually performed via the isobar model or sum of relativistic Breit-Wigner terms representing the different possible implied resonances plus a non-resonant background: can one go beyond?

Section~\ref{FSI} is devoted to some final state interaction (FSI) studies, section~\ref{diagrammatifDM1M2} to a diagrammatic approach and flavor symmetry for $D \to V P$ decays
($V, P \equiv$ vector,  pseudoscalar mesons), section~\ref{FAPDm1m2m3} to a quasi-two-body QCD factorization model for $D$ decays into three mesons and 
section~\ref{Conludingremarks} to amplitude parametrizations based on quasi-two-body factorization  
and to some conclusions.

\section{Final-state interaction constraints}
\label{FSI}
\subsection{Different effective hadronic formalism approaches}
\label{effective_approaches}

The $ K^- \pi^+$ FSI in the $D^+ \to K^- \pi^+ \pi^+$ data of the  E791 Collaboration has been the subject of many studies using different effective hadronic formalism approaches~\cite{Brasil13a, Brasil13b, Patricia15, KubisJHEP1510, Nakamura16}.
In Ref.~\cite{Brasil13a} it is shown that final state interactions are important in shaping the Dalitz plot
and that several weak and hadronic processes are required.
In Ref.~\cite{Brasil13b} the theoretical treatment of this decay includes a rich dynamic behavior that mix weak and strong interactions in a non trivial way.
The authors of Ref.~\cite{Patricia15} assume the dominance of the weak vector current together with a Chiral effective Lagrangian and phenomenological form factors.
In Ref.~\cite{KubisJHEP1510} a full dispersive Khuri-Treiman formalism is applied, resumming rescattering contributions to all orders using $\pi \pi$ and $\pi K$ phase shifts as input, and fitting subtraction constants to the experimental Dalitz plots from CLEO and FOCUS.
S. X. Nakamura~\cite{Nakamura16} has performed a coupled-channel analysis of pseudo-data generated from the isobar model of the E791 Collaboration. 
The authors of Ref.~\cite{Patricia16} have studied the $D^+ \to K^+ K^- K^+$ process with a multi-meson model as an alternative to isobar model, with free parameters predicted by the theory to be fine-tuned by a fit to data.

\subsection{Basic weak interaction plus Chiral unitary approach}

The study of $f_0(980)$ production in $D_s^+ \to \pi^+ \pi^+ \pi^-$ and $D_s^+ \to \pi^+ K^+ K^- $ decays 
performed by the authors  of Ref.~\cite{Oset16} is described below.
 For the sake of completeness the main steps of their derivation is reproduced here.
 They start from the Cabibbo favored $c \to s \bar d u$  flavor changing process. Then the $c \bar s$ pair of the  $D_s^+$ decays into a $\bar d u $ (which hadronizes into a $\pi^+$)
and  a $s \bar s$ pair. 
Insertion,  in the $s \bar s$ pair, of a $q\bar{q}$ with the quantum numbers of the  vacuum, $\bar{u}u+\bar{d}d+\bar{s}s$, leads then, via hadronization, to the production of two pseudoscalar mesons.
This dominant process is depicted in Fig.~1 of Ref.~\cite{Oset16}.
To find out the meson-meson components in the $s\bar s$ pair one can define the following $q\bar{q}$ $M$ matrix:

\begin{equation}
\label{M}
M=\left(
           \begin{array}{ccc}
             u\bar u & u\bar d  & u\bar s  \\
             d\bar u & d\bar d  & d\bar s  \\
             s\bar u & s\bar d  & s\bar s 
           \end{array}
                    \right), 
\end{equation}
which satisfies

\begin{equation}
\label{MdotM}
 M \cdot M = M \times ( \bar u u + \bar d d + \bar s s)
\end{equation}
 In the standard 
$\eta-\eta^\prime$ mixing~\cite{BramonPLB283} the matrix $M$ is related to~\cite{GamermannEPJA41}

\begin{equation}
\label{phi}
\Phi=\left(
  \begin{array}{ccc}
    \frac{1}{\sqrt{2}} \pi^0 + \frac{1}{\sqrt{3}} \eta + 
    \frac{1}{\sqrt{6}} \eta ^{\prime}
    & \pi^+  & K^+ \\
    \pi^- &  - \frac{1}{\sqrt{2}} \pi^0 + \frac{1}{\sqrt{3}} \eta + \frac{1}{\sqrt{6}} 
\eta ^{\prime} & K^0 \\
    K^-       & \bar K^0 &  - \frac{1}{\sqrt{3}} \eta + \sqrt{\frac{2}{3}} \eta ^{\prime}  
           \end{array}
         \right).
\end{equation}
Neglecting  $\eta^\prime$ contribution, its mass being too large, one has:

\begin{equation}
s\bar s  \,  (\bar{u} u  + \bar{d} d + \bar{s} s  )  \equiv  \left( \Phi \cdot \Phi 
\right)_{33} = K^- K^+ + \bar{K}^0 K^0 + \frac{1}{3} \eta\eta, 
\end{equation}
which are the produced states before FSI. 
After rescattering the  $K^+K^-$ pair can produce $\pi^+ \pi^-$ and/or $K^+ K^-$ pairs. 
The $D_s^+$ decay width into a $\pi^+$ and two mesons, labelled, $\Gamma_{P^+P^-}$, where
 $P^+P^- \equiv $ $K^+ K^-$ or $\pi^+ \pi^-$, satisfies

\begin{equation}
\frac{d\Gamma_{P^+P^-}}{dM_{inv}}=\frac{1}{(2\pi)^3}\frac{p_{\pi}\tilde{p}_P}{4M^2_{D_s}}
|T_{P^+P^-} |^2\, ,
\label{decay1}
\end{equation}
with,

\begin{equation}
T_{K^+K^-}=V_0~\left(1+ G_{K^+ K^-}\, t_{K^+ K^- \to K^+ K^-} +
G_{K^0 \bar K^0}\, t_{K^0 \bar K^0 \to K^+ K^-}  +
\frac{2}{3}\,\frac{1}{\sqrt{2}}\,G_{\eta \eta}\, {t}_{\eta \eta \to K^+K^-} \right),
\end{equation}

\begin{equation}
T_{\pi^+\pi^-}=V_0~\left( G_{K^+ K^-}\, t_{K^+ K^- \to \pi^+ \pi^-} +
G_{K^0 \bar K^0}\, t_{K^0 \bar K^0 \to \pi^+ \pi^-}  +
\frac{2}{3}\,\frac{1}{\sqrt{2}}\,G_{\eta \eta}\, {t}_{\eta \eta \to \pi^+\pi^-} \right).
\end{equation}
The function $G_{l}$ is the loop function,
\begin{equation}
G_{l} (s) = i \int\frac{d^{4}q}{(2\pi)^{4}}\frac{1}{(p-q)^{2}-m^2_{1}+i\varepsilon}\,\frac{1}
{q^{2}-m^2_{2}+i\varepsilon},
\end{equation}
$m_1$, $m_2$ being the meson masses in loop $l$.
The integral on $q^0$ is  analytical and a  cut-off, $|{\bf q}_{max}| 
= 600~ \rm{MeV}/c$ is needed in the integral on ${\bf q}$ in order to reproduce the experimental amplitudes.
The $t_{i\to j}$ matrices are obtained by solving the coupled-channel Bethe-Salpeter equation

\begin{equation}
t_{i\to j} (s) = V_{i j} (s) + \sum^{5} _{l=1} V_{i l} (s) G_{l} (s) t_{l\to j} (s),
\end{equation}
where the indices $i, j, l$ running from 1 to 5 denote the different channels: 1 for $\pi^+ \pi^-$, 2 for $\pi^0 \pi^0$, 3 for $K^+ K^-$, 4 for $K^0 \bar K^0$, and 5 for $\eta \eta$.
 The kernel $V_{i j}$ are the tree-level transition amplitudes built from phenomenological 
Lagrangians in Ref.~\cite{Oset87}.

Adjusting $V_0$ and comparing  theoretical amplitudes $T_{K^+K^-}$ and $T_{\pi^+\pi^-}$ with those available from the experimental data~\cite{delAmoSanchez:2010yp, Aubert09} leads to a fair agreement as can be seen in Figs.~2, 3 (and 4) of Ref.~\cite{Oset87}.
Invariant mass distributions for  $D^+_s\to \pi^+\pi^-\pi^+$  and $D^+_s\to \pi^+K^-K^+$ are shown in Fig.~5 of  Ref.~\cite{Oset87}
The $f_0(980)$ signals in the spectra are (up to a global common normalization factor) predictions of the Chiral Unitary approach with no free parameters.
The mechanism displayed above was used in Ref.~\cite{OsetPLB2015} to obtain branching ratios for  $a_0(980)$ and $f_0(980)$ production in good agreement with experiment.
An interesting issue will be the  study of the $\pi^+\pi^0 \eta$ decay mode which
generates the $a_0(980)$ and can lead to informations on possible $f_0(980)$ and $a_0(980)$ mixing.

\section{Diagrammatic approach for $D$ decays into a vector and a pseudoscalar meson} 
\label{diagrammatifDM1M2}
\subsection{$D\to VP$ decays within SU(3) flavor symmetry}
\label{DVP}

The $c$ quark mass, $m_c$, being too high to apply Chiral perturbation theory and too light to use heavy quark expansion approaches, one can use a diagrammatic approach with flavor-flow diagrams classified according to the topologies of weak interactions with all strong interaction effects included. 
This model-independent analysis, based on flavor SU(3) symmetry, determines topological amplitudes allowing to specify the relative importance of different underlying decay mechanisms.
In such an approach, introduced by L.~L.~Chau~\cite{Chau83}, the conceivable topologies of weak interactions are shown in Fig.~1 of Ref.~\cite{ChengPRD85}.

We describe here some results of the recent work of Ref.~\cite{ChengPRD93} in which all
two-body charmed meson decays $D\to VP$ are studied in this diagrammatic framework.
There, within SU(3) flavor symmetry, only four types of amplitudes exist: 
color-allowed $T$, color-suppressed $C$, $W$-exchange $E$, and $W$-annihilation $A$.
Subscript $P$ or $V$ to each amplitude, {\it e.g.}, $T_{P(V)}$, denote the amplitude in which the spectator quark goes to the pseudoscalar or vector meson in the final state. These two kinds of amplitudes do not have {\it a priori} any obvious relationship.
All the flavor amplitude magnitudes and their associated strong phases are then extracted through a fit on existing experimental branching fractions.

\subsection{Fit on branching fractions of  $D\to VP$ decays}
\label{FitBrDVP}
 
The 8 complex amplitudes, $T_{P(V)}, C_{P(V)}, E_{P(V)}, A_{P(V)}$  (15 real parameters, $T_V$ chosen to be real) are determined by performing a  $\chi^2$ fit of 16 experimental branching fractions for Cabibbo-favored $D^0$ and $D_{(s)}^+$ decays proportional to the Cabibbo-Kobayashi-Maskawa (CKM) factors $V_{cs}^*V_{ud}~\sim~{\cal O}(1)$. 
All the data are extracted from the Particle Data Group~\cite{PDG14} but the branching fraction ${\cal B}_{\rho^+\,\eta\,'}$ is taken from Ref.~\cite{PLB750}.
The determination of $D_s^+\to\pi^+\rho^0$ using results from Refs.~\cite{Aubert09, PDG14} (see Ref.~\cite{ChengPRD93}) allows the extraction of the $A_{P,(V)}$ amplitudes.
Among the several solutions found by the authors there is one favored, named (A1).
The results of the fits are shown in the Table II of Ref.~\cite{ChengPRD93}.
Comparison is made in particular with the pole model of Ref.~\cite{YWL} which is built using generalized factorization and addition of poles in annihilation diagrams.

The Cabibbo-favored amplitudes, resulting from the branching fraction fit results (solution A1) are shown in Table~\ref{CFAMP}.

\begin{table}[tp!]
\begin{center}
\begin{tabular}{l c c c c c c c }
\hline
\hline
         $|T_P|$            &$\delta_{T_P}$          &$|C_V|$                          &$\delta_{C_V}$          &$|C_P|$                  &$\delta_{C_P}$                &$|E_V|$                           \\
\hline
$8.46^{+0.22}_{-0.25}$&$57^{+35}_{-41}$&$4.09^{+0.16}_{-0.25}$&$-145^{+29}_{-39}$&$4.08^{+0.37}_{-0.36}$&$-157\pm2$    &$1.19^{+0.64}_{-0.46}$ \\
            \hline
\hline
       $\delta_{E_V}$   &   $|E_P|$                            &$\delta_{E_P}$   &$|A_P|$                           &$\delta_{A_P}$             &$|A_V|$ &$\delta_{A_V}$ \\
\hline
    $-85^{+42}_{-39}$ &     $3.06\pm0.09$&$98\pm5$     &$0.64^{+0.14}_{-0.27}$&$152^{+48}_{-50}$          &$0.52^{+0.24}_{-0.19}$  &$122^{+70}_{-42}$  \\
           \hline
           \hline
\end{tabular}
\caption{Cabibbo-favored amplitudes resulting from the branching fraction fit results~\cite{ChengPRD93} (solution A1) for a $\eta$-$\eta'$ mixing angle of $43.5^\circ$.  
Units: $10^{-6}$, strong phases in degrees. $|T_V|=4.21^{+0.18}_{-0.19}$. 
}
\label{CFAMP}
\end{center}
\end{table}
The modulus of color-allowed tree  $T_P $ amplitude is the  largest. 
The moduli  of  color-allowed tree $T_V $, color-suppressed tree $C_{V(P)}$ and $W$-exchange $E_P$ are of the same magnitude.
The moduli of the $W$-annihilation   $A_{P(V)}$ amplitudes are the  smallest.

Branching fraction predictions, with no flavor  SU(3) breaking, for singly Cabibbo-suppressed decays proportional to $V_{cd}^*V_{ud}\sim {\cal O}(\lambda)$ and  to $V_{cs}^*V_{us}\sim {\cal O}(\lambda)$ with $\lambda = 0.22543$ (CKMfitter, see Ref.~\cite{CKMfit}) can be seen in Table III of Ref.~\cite{ChengPRD93} and predictions for doubly Cabibbo-suppressed decays (no SU(3) breaking)  proportional to $V_{cd}^*V_{us}\sim {\cal O}(\lambda^2)$ in Table IV.
The predictions for the doubly Cabibbo-suppressed channels are in good agreement with data but
the singly Cabibbo-suppressed ones have some flavor SU(3) symmetry breaking effects. 

\subsection{Concluding remarks on this $D\to VP$ study}
\label{Conclu_DVP}

Exact flavor SU(3) describes reasonably well the available data.
If $T$ and $C$  amplitudes are factorizable, the  effective Wilson coefficients $a_{1,2}$, $|a_2/a_1|$ and arg$(a_2/a_1)$ (see next section) can be extracted from  Cabibbo-favored $D^+\to\ol{K}^{*0}\,\pi^+$ and $\ol{K}^0\,\rho^+$ [solution (A1)]. 
 The results are shown in Table~VII of Ref.~\cite{ChengPRD93}.
 SU(3) symmetry breaking in color-allowed  $T$ and color-suppressed  $C$ tree amplitudes is needed in general to have a better agreement with experiment.
 Nevertheless, the exact flavor SU(3)-symmetric approach alone is adequate to provide an overall explanation for the current data.
The impact of this symmetry on $D \to PP$ decays has been presented in this workshop by P. Santorelli~\cite{SantorelliCPV}.
Within the diagrammatic approach one should quote the validity study of flavor SU(3)~\cite{RosnerPRD79}.
 
\section{Factorization approach for hadronic three-body $D$ decays} 
 \label{FAPDm1m2m3}

\subsection{  Quasi-two-body factorization for  $D^0 \to K_S^0 \pi^+ \pi^-$}
\label{D0K0PP}

QCD factorization beyond na\"ive factorization, expansion in $\alpha_s$ (strong coupling constant) and $1/m_b$ ($m_b$ $b$-quark mass), applies with success to charmless nonleptonic two-body  $B$ decays~\cite{BenekeNPB675}.
In $D$ decays,  $m_c \sim m_b/3$  leads a priori to significant corrections to the factorized results and
factorization is more a phenomenological approach, based on the seminal work by Bauer, Stech and Wirbel~\cite{BauerZPC34}.
It is then applied successfully to $D$ decays, treating  Wilson coefficients as phenomenological parameters to account for non-factorizable corrections.
A part from a recent attempt to extend the framework of QCD factorization to non-leptonic
$B$ decays into three light mesons~\cite{KMV2015},
so far there exists no factorization theorem for  three-body decays.
However there are important contributions from intermediate resonances as
$\rho(770)$, $K^*(892)$ and $\phi(1020)$  and  three-body decays may be
considered as quasi-two-body decays. 
 One makes the hypothesis that two of the three final-state mesons form a single state originating from a
quark-antiquark pair. This  leads to a quasi-two-body final state to which the  factorization procedure is applied.

Within this framework we report on the  Dalitz plot studies of  $D^0 \to K_S^0 \pi^+ \pi^-$ decays performed in Ref.~\cite{JPD_PRD89}.
There is  no penguin ($W$-loop diagram) in  this decay and the weak effective Hamiltonian reads

 \begin{equation}
 H_{eff}=\frac{G_F}{\sqrt{2}} V_{CKM} \sum_{i=1,2} C_i(\mu) O_i(\mu) + h.c.,
 \end{equation}
 where $V_{CKM}$ represents the  quark mixing couplings,
$G_F=1.166\times 10^{-5}$ GeV$^{-2}$  the Fermi coupling, $C_i(\mu)$  the QCD Wilson coefficients arising from $W$  exchange and  $\mu$ the renormalization scale with $ \mu \sim  m_c=1.3$ GeV. 
The left-handed quark current-current operators $O_{1}$ reads

  \begin{equation}
   O_1=j_1 \otimes j_2,  
  j_1 =\bar s_\alpha\gamma^\nu(1-\gamma^5)c_\alpha\equiv  (\bar s c)_{V-A}, 
   j_2= \bar u_\beta\gamma_\nu(1-\gamma_5)d_\beta\equiv (\bar u d)_{V-A}, 
 \end{equation}
$\alpha$ and $ \beta$ being color indices.
The operator $O_{2}$ has a similar expression. 

 In the amplitude and at leading order in $\alpha_s$, the following real effective  QCD  coefficients $a_1(m_c)$ and $a_2(m_c)$ will appear,

  \begin{equation}
 a_1(m_c)=C_1(m_c)+\frac{C_{2}(m_c)}{N_C},\hspace{1cm} a_2(m_c)=C_2(m_c)+\frac{C_{1}(m_c)}{N_C},
 \end{equation}
$N_C=3$ being the number of colors (from now on $a_i(m_c)\equiv a_i, i=1,2$).
The use of the  Operator Product Expansion and of the fact that the $W$  mass is large  leads to the two-body factorization approximation,

  \begin{equation}
\label{factorization}
\langle M_1M_2\vert O_i(\mu)\vert D^0\rangle = \langle M_1\vert j_1\vert 0\rangle
 \langle M_2\vert j_2\vert D^0\rangle + {\rm higher\  order\ corrections}.
 \end{equation}
 
  \begin{figure}[t]
 \begin{center}
\includegraphics[scale=0.45]{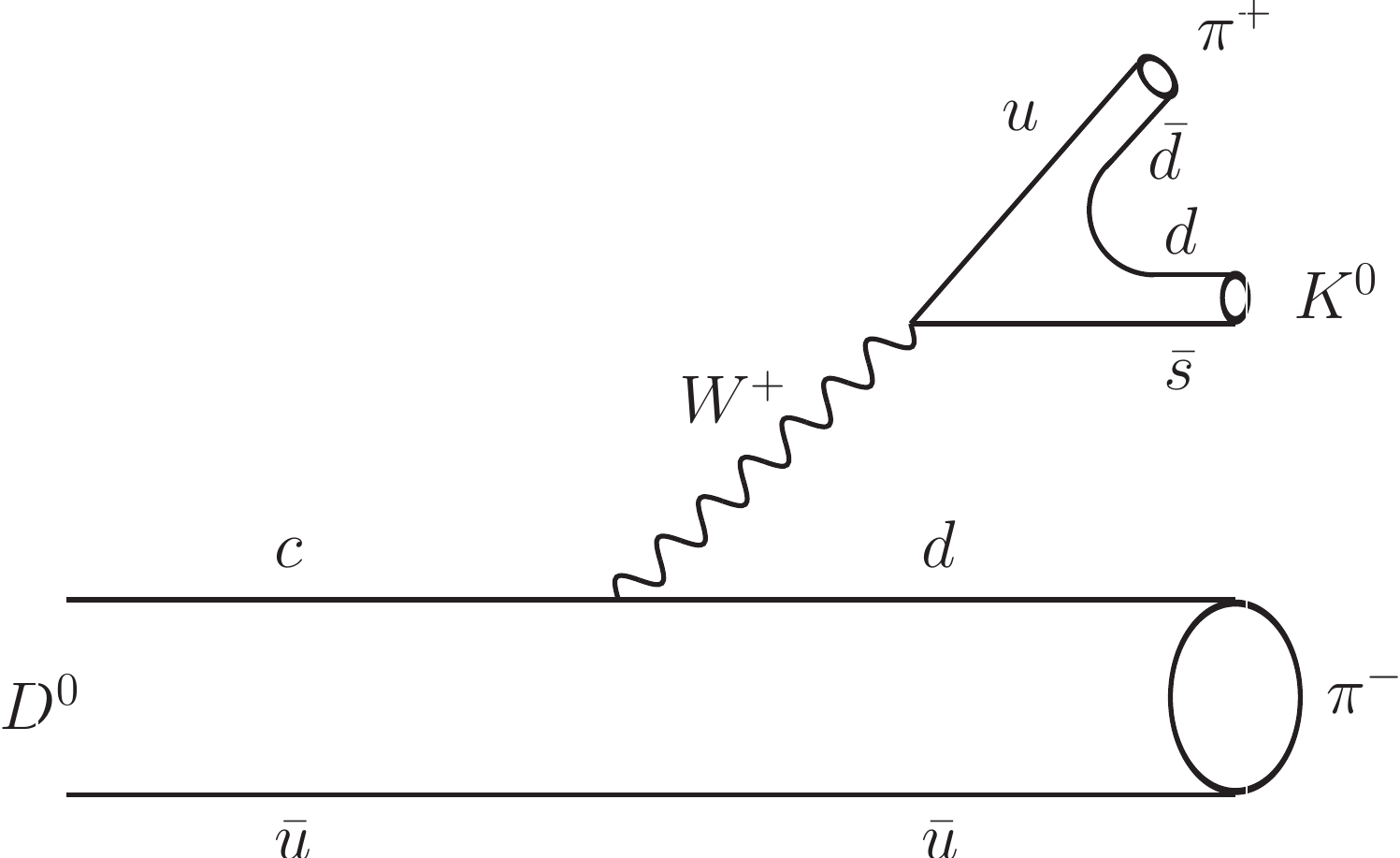}
\hspace{0.2cm}
\includegraphics[scale=0.45]{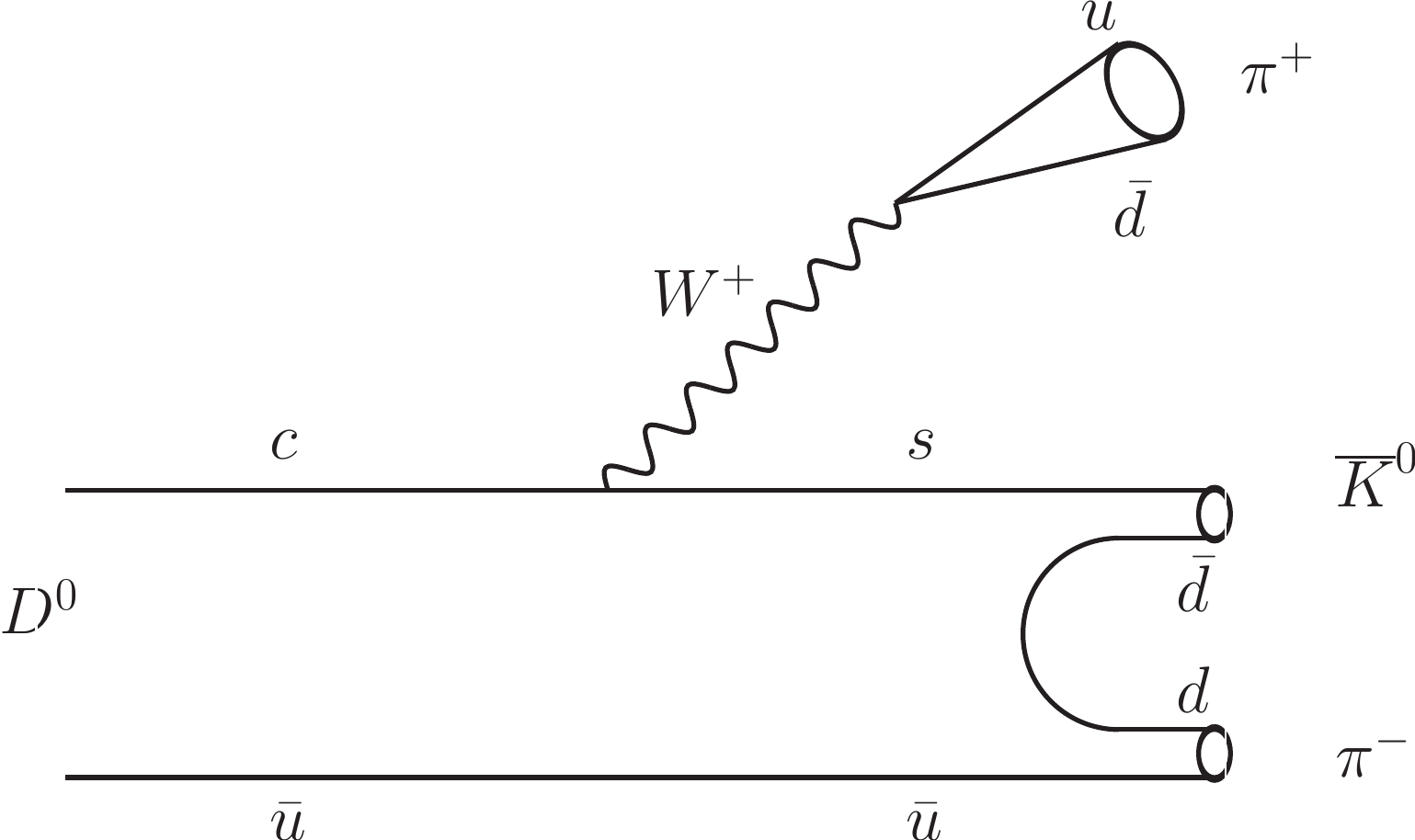}
\caption{Microscopic quark tree diagrams for doubly Cabibbo-suppressed (left) and Cabibbo-favored (right) amplitudes}
\label{quark_diagrams}
\end{center}
\end{figure}

In the quasi-two-body approximation, 
the final state $\bar K^0 \pi^+ \pi^-$ is assimilated to the two-body states $[\bar K^0 \pi^\pm]_{L}\ {\pi^\mp}$ and $\bar K^0\  [\pi^+{\pi^-}]_{L}$ where the meson-meson pair can be in $L= S,P$ and $D$ states.
The mesons $M_{1,2}$ of the factorization equation~(\ref{factorization}) are then  
$M_1=[\pi^+K^0]_L, M_2=\pi^-$  or $M_1= \bar K^0, M_2= [\pi^+ \pi^-]_L$.  
In the $D^0 \to K_S^0 \pi^+ \pi^-$ decays the possible $c$ quark flavor changing are $c \to su\bar d$ leading to Cabibbo-favored (CF) amplitudes $\propto$ $V_{CKM}=V^*_{cs}V_{ud}\equiv \Lambda_1$ and $c \to du\bar s$ for doubly Cabibbo-suppressed (DCS) amplitudes $\propto$ $V_{CKM}=V^*_{cd}V_{us}\equiv \Lambda_2$.
Here, besides tree amplitudes, there are annihilation ones arising from $W$ exchange between the quarks $c$ and $\bar u$ of the $D^0$.
The interested reader will find the detailed expressions of the different amplitudes obtained applying the above quasi-two-body factorization approximation to the 
$\left \langle \bar K^0 \ \pi^- \pi^+ \vert \ H_{eff}\ \vert D^0 \right \rangle$ matrix element.
Let us exemplify some terms entering these amplitudes. In the DCS tree amplitude (see Eq.~(10) of Ref.~\cite{JPD_PRD89}) for $L=S$ one encounters the contribution of a term like (see left diagram of Fig.~\ref{quark_diagrams})
 
\begin{equation}
\frac{G_F}{2}\ \Lambda_2   a_1  \langle \pi^-\vert (\overline  d \ c)_{V-A}\vert D^0 \rangle 
\cdot \ \langle [K^0 \pi^+ ]_S\vert (\bar  u \ s)_{V-A}\vert 0   \rangle, 
\end{equation} 
where the first matrix element is related to 
the $D^0 \to \pi$ transition form factor and the second one to the scalar $K \pi$ form factor.
In the CF tree amplitude (see Eq.~(6) of Ref.~\cite{JPD_PRD89}) for $L=S$ one has the contribution of a term as (see right diagram of Fig.~\ref{quark_diagrams})

\begin{equation} 
\frac{G_F}{2}\ \Lambda_1  a_1  \langle \pi^+\vert (\bar  u \ d)_{V-A}\vert 0  \rangle 
\cdot \langle  [\bar K^0 (p_0){\pi^-}]_S \vert (\overline  s \ c)_{V-A}\vert D^0(p_{D^0})\rangle,
\end{equation}
where the first matrix element is related to 
the $\pi$ decay constant and the second one to the scalar $D^0 \to K \pi$ transition form factor.
Its evaluation is less straightforward, it could be experimentally evaluated from semi-leptonic processes such as, $D^0 \to K^- \pi^+ \mu^+ \mu^-$~\cite{D0semilep, Mariana}.
Here assuming this transition to proceed through the dominant intermediate resonance $K^*_0(1430)$, it can be written in terms of the $K \pi$ scalar form factor.
The corresponding amplitude given by Eq.~(13) of Ref.~\cite{JPD_PRD89} is

\begin{equation} 
\label{A1S}
T^{CF}_{[\overline{K}^0 \pi^-]_S \ \pi^+}(s_0, s_-,s_+) 
= -\frac{G_F}{2}\ a_1 \Lambda_1\ \chi_1 \ \left (m_{D^0}^2-s_-\right ) \ f_\pi \ F_0^{D^0 K^*_0(1430)^-}(m_\pi^2)\  F_0^{\overline{K}^0\pi^-}(s_-),
\end{equation}
where $s_\pm=(p_{\pi^\pm}+p_{K^0})^2=m_\pm^2, s_0=(p_{\pi^+}+p_{\pi^-})^2=m_0^2$. 
The scalar $K \pi$ form factor $F_0^{\overline{K}^0\pi^-}(s_-)$ includes the contributions of the $K^*_0(800)$ and $K^*_0(1430)$ resonances.
The factor $\chi_1$, related to the strength of this scalar form factor is taken as a complex constant to be fitted.
It can be estimated from the $K_0(1430)$ properties~\cite{JPD_PRD89}.
Here $f_\pi$ is the pion decay constant.
Following Ref.~\cite{ChengPRD81_074031} the value of 0.48 is used for the $D^0$ to $K^*_0(1430)$ transition form factor $F_0^{D^0 K^*_0(1430)^-}(m_\pi^2)$.

\begin{figure}[h] 
\begin{center}
\includegraphics[scale = 0.35]{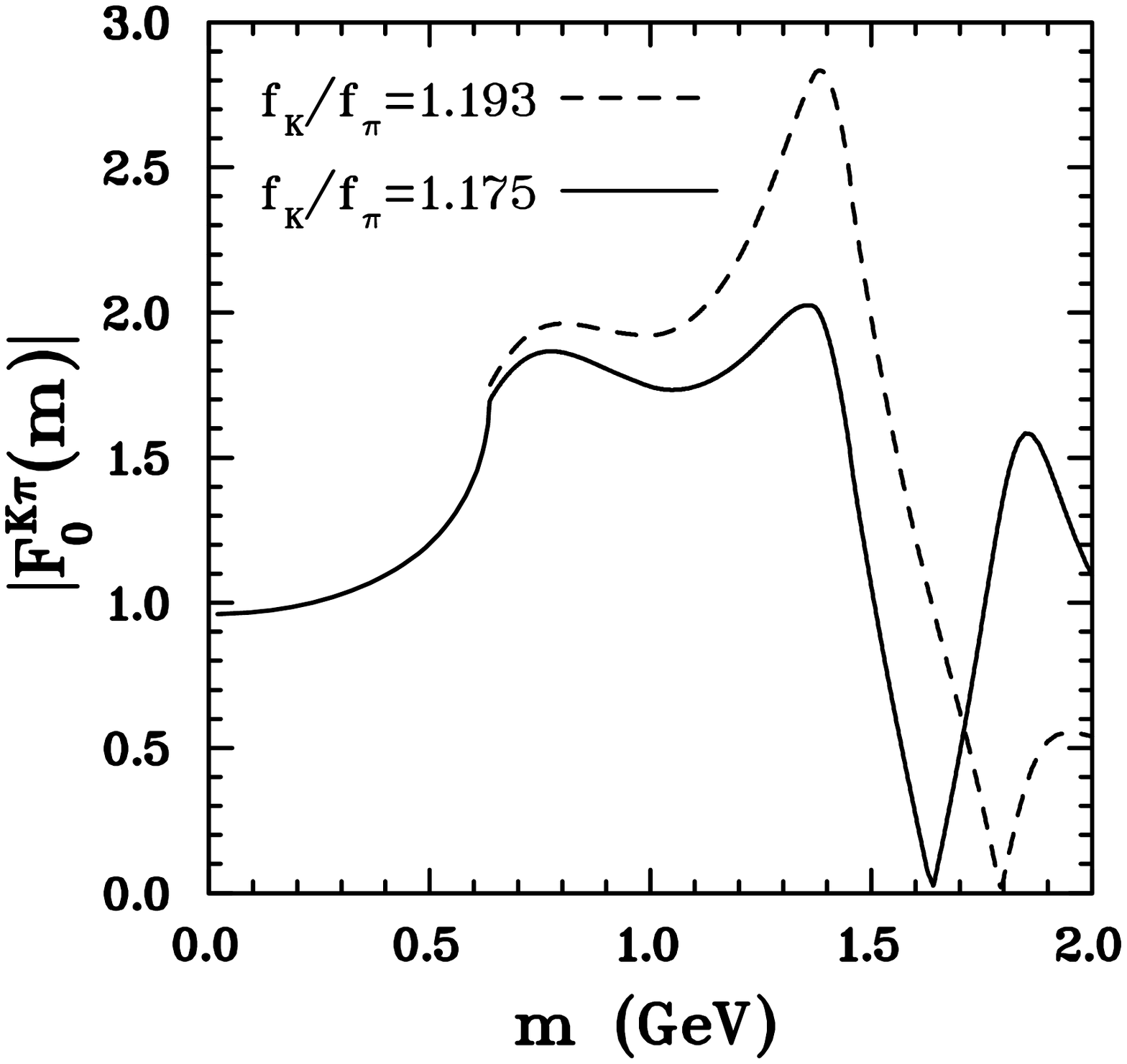}~~~~
\includegraphics[scale = 0.35]{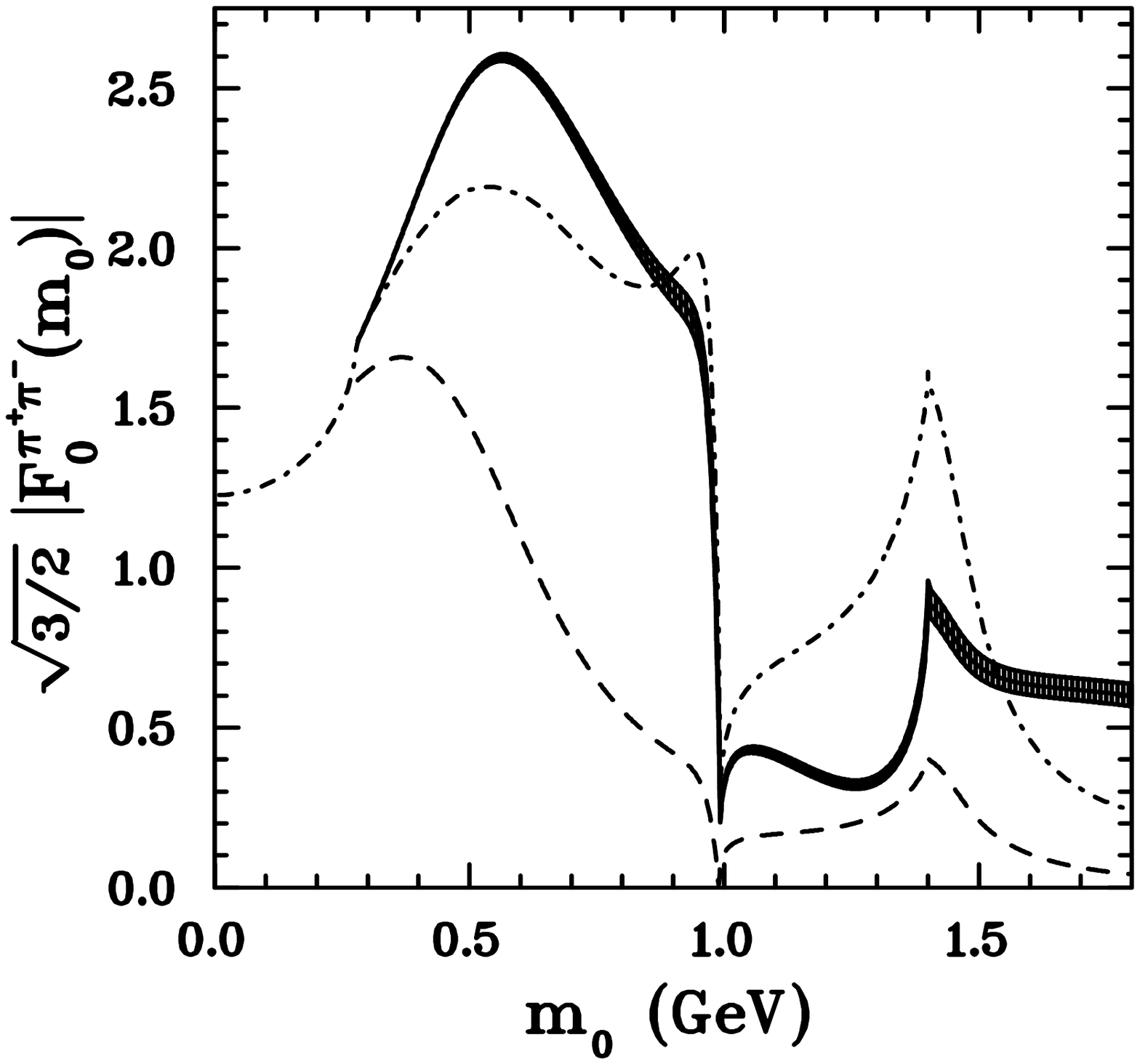}~~~~
\vspace{0.2cm}
\caption{Left panel: modulus of the scalar $K \pi$, $F^{K\pi}_0$, form factor as a function of the effective $K \pi$ mass for two values of the $f_K/f_\pi$ ratio ($f_K$ being the kaon decay constant).
Right panel: modulus of the scalar $\pi \pi$, $F^{\pi \pi}_0$, form factor as a function of the effective
 $\pi \pi$ mass, dark band variation within the parameter errors of the fit to Belle data~\cite{A.Poluektov_PRD81_112002_Belle}, dashed line same form factor but different parameters (Ref.~\cite{DedonderPol}), dot-dash line is the result of Ref.~\cite{Moussallam_2000} using Muskhelishvili-Omn\`es equations.
}
\label{F0Kpi_pipiff}
\end{center}
\end{figure}
 
What we have shown just above for amplitudes with $K \pi$ final state pair in $S$  wave, works also for
the $P$-wave case and for amplitudes with the $\pi^+ \pi^-$  final state pair in $S$ and $P$ wave.
So, amplitudes with the $\pi^+ \pi^-$ ($K \pi$) final state pair in $S$ and $P$  wave are described in terms of scalar and vector $\pi \pi$ ($K \pi$) form factors.
Form factors with final meson pair in $D$ wave are represented by relativistic Breit-Wigner formulae.
The 13~tree-amplitudes and 14~$W$-exchange ones derived in Ref.~\cite{JPD_PRD89} can be recast into 10 amplitudes.
Summary of these CF and DCS, amplitudes associated to the different quasi two-body channel together with the contributing dominant resonances are listed in Table 1 of Ref.~\cite{JPD_PRD89}.

\subsection{Form factors}

It can be shown from field theory and using dispersion relations~\cite{Barton65} that meson-meson form factors can be calculated exactly, using Muskhelishvili-Omn\`es 
equations, if one knows the meson-meson strong interactions at all energies. 
The details and  corresponding references of the unitary scalar and vector $K\pi$ and  $\pi \pi$ form factors used in the best fits of the  $D^0 \to K^0_S \pi^+ \pi^-$  Dalitz plot performed in Ref.~\cite{JPD_PRD89} can be found in this reference.
We just plot here, in Fig.~\ref{F0Kpi_pipiff}, the modulus of the scalar $K\pi$ ($F^{K\pi}_0$, left panel) and $\pi \pi$ ($F^{\pi \pi}_0$, right panel) form factors as function of the
$K\pi$ ($m_\pm$) and $\pi \pi$ ($m_0$) effective masses.  
The scalar $K\pi$ form factor is characterized by two bumps arising from the $K_0^*(800)$ and $K_0^*(1430)$ contributions and the $\pi \pi$ form factor by a dip coming from the $f_0(980)$) and two bumps from the $f_0(500)$ and $f_0(1400)$.

\subsection{Dalitz-plot fit}

The Dalitz plot distribution of the best fit to the Belle data~\cite{A.Poluektov_PRD81_112002_Belle} is shown in Figs.~9 of Ref.~\cite{JPD_PRD89}.
This distribution reproduces well that of Belle.
The corresponding input parameters and the obtained 33 free parameters are given in section IV and in Table II of Ref.~\cite{JPD_PRD89}, respectively.
The Dalitz plot shows a rich interference pattern with the dominance of the $K^*(892)$. 

Branching fraction (Br), listed in Table V of Ref.~\cite{JPD_PRD89}, compare well with those of Belle's analysis.
Their sum, equal to  133~\%,  shows the importance of interferences.
The largest Br come from the amplitudes, 
$\mathcal{M}_{1}$ [contributions of $K^*_0(800)^-$, $K^*_0(1430)^-$]  with a Br of 25.~\%) ,  $\mathcal{ M}_{2}$  [contribution of   $f_0(500)$, $f_0(980)$, $f_0(1400)$] with a Br of 16.9~\%,  $\mathcal{M}_{3}$  [contributions of $K^*(892)^-$] with a Br of 62.7~\%) and $\mathcal{M}_{4}$ [contributions of $\rho(770)^0$] with a Br of~21.9~\%.
The sum of the Belle Br with the final $\pi^+ \pi^-$ pair state in $S$ wave, equal to 18.6\%, is close the Br of $\mathcal{ M}_{2}$, equal to 16.9~\%.
The annihilation ($W$ exchange) contributions can be important.

\section{Concluding remarks}
\label{Conludingremarks}
\subsection{Amplitude parametrizations}
\label{AP}

 As a sound {alternative} to the simplistic and widely used 
{ isobar model}, explicit amplitude  parametrizations, that can be readily implemented in experimental analysis, are suggested in Refs.~\cite{Boito_LHCb15, BDE2KL2}  for the study of the decays
 { $D^+ \to \pi^- \pi^+ \pi^+ $},
{ $D^+ \to K^- \pi^+ \ \pi^+ $}, { $D^0 \to K^0_S \ \pi^+\ \pi^- $  and
{ $D^0 \to K^0_S \ K^+ K^-$}.
These parametrizations, where two-body hadronic final state interactions are
taken into account in terms of { unitary} $S$- and $P$-wave $\pi\pi$,
$\pi K$ and $K \bar K$ {form factors},
are derived from different phenomenologically successful works (except for the $D^0 \to K^0_S \ K^+ K^-$ not tested yet) based on quasi two-body factorization approach.

\subsection{Outlook}

\subsubsection{Some other studies on hadronic $D$ decays}
Within the factorization approximation and taking into account final state interaction, two-body hadronic $D^0$, $D^+$ and $D_s^+$ decays have been studied in Ref.~\cite{Bucella95}.
A reasonable agreement with the data, which show some large flavor SU(3) symmetry violation, is obtained.
CP violating asymmetries are also discussed.

The authors of Ref.~\cite{BIGI14} explore consequences of  { constraints} from CPT symmetry on three-body $D$ decays.
They simulate the $D^\pm \to  \pi^\mp K^+ K^-$ decays  and discuss { correlations} with measured { $D^\pm \to  \pi^\mp \pi^+ \pi^-$}.

The impact of New Dynamics is studied in Ref.~\cite{Ikaros15}.

 The authors of Ref.~\cite{ZouCPC37} perform an analysis of pure { annihilation} type  diagrams for two-body $ D\to PP(V)$ decays based on the { $k_T$ factorization}.
Their results agree with the existing experimental data for most channels.

\subsubsection{Concluding summary}

There is and there will be an impressive amount  of {high-quality hadronic-multibody decay data} of { $D^0$, $D^+$, $D_s^+$}.
Improved models are and will be needed for extracting accurate informations from these data.
In this brief review I described some { available potentialities} for constraining  amplitude analyses in some of these charm decays.
I have listed different effective hadronic formalism approaches.
Detailed outcomes of a model, with combination of basic elements of weak interaction together with final state constraints in the framework of a Chiral unitary approach in coupled channel, have been reported.

I have described the diagrammatic-approach framework.
It consists of a model-independent analysis, based on { flavor SU(3) symmetry}  topological amplitudes, allowing to understand  the relative importance of different underlying decay mechanisms.
In this framework a global analysis of two-body $D$ decays into a vector meson and a pseudoscalar meson has been presented.

A quasi-two-body QCD factorization model for $D$ decays into three mesons and its recent application to $D^0 \to K_S^0 \pi^+ \pi^-$ was shown to describe successfully the data.
It suggests to go beyond the superposition of Breit-Wigner amplitudes in the analysis
of three-body decays with pions and kaons in the final state.
It is advocated to use  a phenomenological  quasi-two-body factorization
approach in which two-body hadronic final state interactions are fully
taken into account in terms of unitary $S$- and $P$-wave $\pi\pi$, $\pi K$ and $K \bar K$ 
form factors}.

It could be interesting to see if the above phenomenological  quasi-two-body factorization approach could be used in amplitude analysis of four-body hadronic $D$ decays such as $D^0 \to K^+ K^- \pi^+ \pi^-$ recently analyzed by the CLEO Collaboration~\cite{CLEOPRD85},   $D^0 \to \pi^+ \pi^- \pi^+ \pi^-$~\cite{DargentD4pi} and $D^0 \to K^- \pi^+ \pi^+ \pi^-$ presently studied  by the BES III Collaboration~\cite{YuLu}.

\vspace{0.8cm}
\noindent
Acknowledgments
\\
I thank Jean-Pierre Dedonder, Robert Kami\'nski and Leonard Le\'sniak for their fruitful collaboration on the quasi-two-body factorization work on the  $D^0 \to K_S^0 \pi^+ \pi^-$ decays (IN2P3-COPIN collaboration agreement, project N$^\circ$ 08-127).
I appreciate helpful discussions and fruitful comments of Jean-Pierre Dedonder during the preparation of my talk and the writing of this contribution.
I am obliged to Diogo Boito and Bruno El-Bennich for their help on the amplitude parametrizations proposed in section~\ref{AP}.

\end{document}